# scientific reports

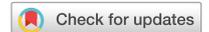
Check for updates

# Field induced crossover in critical behaviour and direct measurement of the magnetocaloric properties of La$_{0.4}$Pr$_{0.3}$Ca$_{0.1}$Sr$_{0.2}$MnO$_3$

Sagar Ghorai[1]✉, Ridha Skini[1], Daniel Hedlund[1], Petter Ström[2] & Peter Svedlindh[1]

La$_{0.4}$Pr$_{0.3}$Ca$_{0.1}$Sr$_{0.2}$MnO$_3$ has been investigated as a potential candidate for room temperature magnetic refrigeration. Results from X-ray powder diffraction reveal an orthorhombic structure with Pnma space group. The electronic and chemical properties have been confirmed by X-ray photoelectron spectroscopy and ion-beam analysis. A second-order paramagnetic to ferromagnetic transition was observed near room temperature (289 K), with a mean-field like critical behaviour at low field and a tricritical mean-field like behaviour at high field. The field induced crossover in critical behaviour is a consequence of the system being close to a first-order magnetic transition in combination with a magnetic field induced suppression of local lattice distortions. The lattice distortions consist of interconnected and weakly distorted pairs of Mn-ions, where each pair shares an electron and a hole, dispersed by large Jahn–Teller distortions at Mn$^{3+}$ lattice sites. A comparatively high value of the isothermal entropy-change (3.08 J/kg-K at 2 T) is observed and the direct measurements of the adiabatic temperature change reveal a temperature change of 1.5 K for a magnetic field change of 1.9 T.

The need for reduced emission of greenhouse gases, emitted from conventional liquid gas refrigeration devices, has resulted in an increasing interest towards solid state refrigeration systems based on the magnetocaloric effect (MCE)[1]. Historically, in 1949, Giauque and MacDougal received the Nobel Prize for their achievement to reach a temperature below 1 K, using the magnetocaloric effect[2]. The next large breakthrough came in the end of the 90's, with the discovery of 'Giant' MCE and the development of a magnetic refrigerator using Gd as refrigerant by Ames laboratory and Astronautics Corporation of America, making the dream of a magnetic refrigerator possible[3].

Since then, several giant magnetocaloric alloy-materials have been proposed as potential refrigerants for magnetic refrigeration[4–8]. Although these alloys show high adiabatic temperature changes ($\Delta T_{ad}$) near room temperature, their corrosive properties and large thermal hysteresis limit their use in magnetic refrigeration. The perovskite manganites (A$_{1-x}$B$_x$MnO$_3$, where A is trivalent rare earth cation and B is divalent alkaline earth cation) have been put forward as candidates as magnetic refrigerants owing to their chemical stability, ease to synthesise, resistance to corrosion, negligible thermal hysteresis, large relative cooling power (RCP) and magnetic phase transitions near room temperature[9–12]. Substantial research work[13] has shown that the key factor for magnetic refrigeration is the adiabatic temperature change which is defined as[14],

$$\Delta T_{ad} = -\mu_0 \int_0^{H_f} \frac{T}{C_{p,H}} \left( \frac{\partial M}{\partial T} \right) dH, \qquad (1)$$

where $\mu_0$ is free-space permeability, $H_f$ is the applied magnetic field, and $C_{p,H}$ is the heat capacity at constant pressure and field. Materials with high saturation magnetic moments and a Curie temperature ($T_C$) near room temperature are desirable for room temperature magnetic refrigeration[15]. To satisfy these two requirements, in our previous work the non-magnetic La$^{3+}$ ion was partially substituted with the magnetic Pr$^{3+}$ ion in the La$_{0.7}$Ca$_{0.1}$Sr$_{0.2}$MnO$_3$ compound and an enhancement of the saturation magnetic moment along with and enhanced

[1]Solid State Physics, Department of Materials Science and Engineering, Uppsala University, Box 35, 751 03 Uppsala, Sweden. [2]Applied Nuclear Physics, Department of Physics and Astronomy, Uppsala University, Box 516, 751 20 Uppsala, Sweden. ✉email: sagar.ghorai@angstrom.uu.se

nature research

isothermal entropy-change and relative cooling power were observed in the compound $La_{0.5}Pr_{0.3}Ca_{0.1}Sr_{0.2}MnO_3$[12]. As a consequence of the substitution, $T_C$ also shifted from 343 to 296 K[16,17].

It has been observed in studies using active magnetic regenerator (AMR) cycles, that a material with a large isothermal entropy-change in a narrow temperature span demonstrates significantly lower cooling power than a material with a moderate isothermal entropy-change in a wide temperature span[13,15]. Hence, research should focus on finding materials with a high $\Delta T_{ad}$ in a broad temperature span to enhance the cooling power in real applications. This is the advantage of oxide materials, their second-order magnetic phase transition results in a broad temperature span in which magnetic cooling is effective. However, to compensate for the high value of the heat capacity of oxide materials, a sufficiently large magnetic field is required to run the AMR cycle. In recent years there has been substantial progress in magnetocaloric research using pulsed magnetic field as high as 60 T[18–22], implying that there is a possibility in future to use the oxides in AMR or similar refrigeration cycles with a high magnetic field.

In this work, we have investigated the structural, electronic, chemical, magnetic, magnetocaloric and universal scaling properties of the $La_{0.4}Pr_{0.3}Ca_{0.1}Sr_{0.2}MnO_3$ (LP3) compound. The isothermal entropy-change as well as the adiabatic temperature change have been measured in the LP3 compound. An estimation of the required magnetic field for the material to be useful in a real AMR cycle is also presented. With the exception of $La_{0.67}Ca_{0.33}MnO_3$, which shows a first-order magnetic transition near 268 K[23], the $\Delta T_{ad}$ value obtained for our studied sample is relatively higher comparing with reported values for other manganite systems and moreover has the advantage of being recorded near room temperature. Interestingly, a field induced change in critical behaviour was observed in this compound. Using critical scaling analysis, we have identified the critical exponents in the low and high field regions. We argue that the mean-field critical behaviour is a consequence of local lattice distortions consisting of weakly distorted pairs of Mn-ions, where each pair shares one electron and one hole, dispersed by Jahn–Teller (JT) distorted $Mn^{3+}$-ion lattice sites[24]. The magnetization process and the magnetic phase transition at low field is driven by long-range interacting spin clusters consisting of interconnected Mn pairs, while the crossover at higher field involves suppression of distortions at JT-distorted lattice sites gradually removing the cluster-like character of the magnetic state.

## Experimental details

The **LP3** compound was prepared by solid-state reaction as described in our previous work[12]. The structural properties of the sample were characterized by X-ray powder diffraction (**XRPD**) at room temperature using Cu-$K_\alpha$ radiation (Bruker D8-advance diffractometer) by steps of 0.012° with a delay time of 10 s per step. The elemental analysis of the **LP3** sample was performed by Rutherford backscattering spectrometry (RBS) with 2 MeV $^4$He$^+$ and 10 MeV $^{12}$C$^{3+}$ beams, as well as by time-of-flight elastic recoil detection analysis (ToF-ERDA)[25] with 36 MeV $^{127}$I$^{8+}$. The beam was hitting the sample at 5° incidence angle with respect to the surface normal for both RBS measurements, and the energy detector was placed at a backscattering angle of 170°. The sample was wiggled within a 2° interval to average out possible channelling effects in the grains. A silicon drift type detector for particle induced X-ray emission (PIXE) was also present at 135°. For ToF-ERDA the incidence angle was 23° ± 1° with respect to the sample surface, and recoils were detected at 45°. X-ray photoelectron spectroscopy (**XPS**) was used to analyse the oxidation state of the sample. A "PHI Quantera II" system with an Al-K$\alpha$ X-ray source and a hemispherical electron energy analyser with a pass energy of 26.00 eV was used to record the XPS spectra. Pre-sputtering with Ar-ions of 200 eV for 12 s was done on the sample before collecting the XPS spectra in order to remove surface impurities without affecting the sample's properties. The magnetic measurements were performed in a Quantum Design MPMS XL in the temperature range from 390 to 5 K with a maximum field of 5 T. The adiabatic temperature change ($\Delta T_{ad}$) was measured in steps of 0.1 T up to 1.90 T in a home-built device at the Technical University of Darmstadt, Functional Materials[26].

## Results and discussion

**Crystal structure.** The XRPD pattern of the **LP3** sample is shown in Fig. 1a. The Rietveld refinement (using the Fullprof program[27]) of the XRPD pattern confirmed the orthorhombic structure with Pnma space group for the sample, similar as previously reported for $La_{0.5}Pr_{0.3}Ca_{0.1}Sr_{0.2}MnO_3$[12]. No other significant phases have been observed from the XRPD pattern. Details of the lattice parameters are listed in Table 1. For a typical rhombohedral structure, the value of the Goldschmidt tolerance factor ($t_G = \frac{r_A + r_O}{\sqrt{2}(r_B + r_O)}$, where $r_A$, $r_B$ and $r_O$ are the ionic radii of A, B and oxygen ions, respectively) is $0.96 < t_G < 1$, while for an orthorhombic structure the value of $t_G$ is $< 0.96$[28]. For the **LP3** compound the calculated value of $t_G$ is 0.92, which confirms the observed orthorhombic structure of the compound. Most importantly, due to the insignificant change (0.07%, 0.06% and 0.18% changes in $a$, $b$ and $c$ lattice parameters, respectively) of unit cell parameters in the **LP3** compound comparing with those reported for $La_{0.5}Pr_{0.3}Ca_{0.1}Sr_{0.2}MnO_3$[12], the results obtained for the **LP3** compound provide the opportunity to single out effects due to an increased $Pr^{3+}$-ion substitution.

**Electronic structure.** In manganites, with different valence states of Mn, different exchange interactions exist; $Mn^{3+}$–$O^{2-}$–$Mn^{4+}$ giving rise to ferromagnetic (**FM**) double-exchange interaction and $Mn^{3+}$–$O^{2-}$–$Mn^{3+}$ (or $Mn^{4+}$–$O^{2-}$–$Mn^{4+}$) giving rise to antiferromagnetic (**AFM**) super-exchange interaction[30]. Thus, the average Mn-oxidation state plays a crucial role in determining the influence of FM and AFM interactions. XPS has been used to determine the oxidation states of the **LP3** compound and the results are shown in Fig. 1b. The observed $Mn^{3+}/Mn^{4+}$ ratio was 2.3(7), which is comparable to the expected value 2.33. The lack of satellite peaks confirms the absence of any $Mn^{2+}$ state[31]. The observed spin–orbit splitting between Mn-$2p_{1/2}$ and Mn-$2p_{3/2}$ was 11.62 eV, which indicates a decrease of the spin–orbit splitting with increasing $Pr^{3+}$ substitution compared to the result for our previously reported compound $La_{0.5}Pr_{0.2}Ca_{0.1}Sr_{0.2}MnO_3$[12].

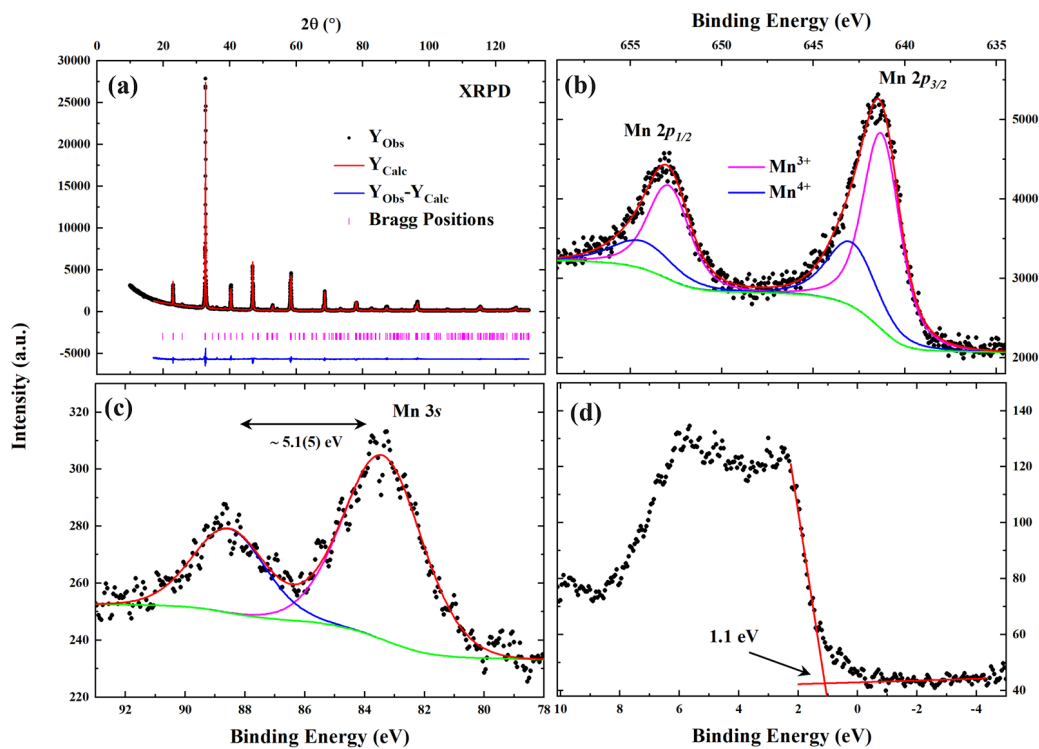

**Figure 1.** (**a**) Rietveld refined XRPD pattern and XPS spectra of (**b**) Mn-2p, (**c**) Mn-3s and (**d**) valence band of the **LP3** compound.

**Table 1.** Lattice parameters and bond information from Rietveld analysis of XRPD pattern.

| Compound | La$_{0.4}$Pr$_{0.3}$Ca$_{0.1}$Sr$_{0.2}$MnO$_3$ |
|---|---|
| Phase | Orthorhombic (Pnma) |
| Lattice parameters (Å) | $a = 5.46124(6)$<br>$b = 7.71813(8)$<br>$c = 5.49213(6)$ |
| Bond lengths (Å) | Mn-O1: 1.9598(13)<br>Mn-O2: 1.952(4) |
| Bond angles (°) | Mn-O1-Mn: 159.95(5)<br>Mn-O2-Mn: 165.4(4) |
| Rietveld refinement parameters[29] | $R_P = 5.56$, $R_{WP} = 7.64$, $R_B = 6.19$ |

The Mn-3s spectrum is shown in Fig. 1c. The parallel and anti-parallel coupling of the spins of 3s core holes and 3d electrons generate two distinct peaks (high-spin and low-spin state at lower and higher binding energy, respectively) in the Mn-3s spectrum[32,33]. Beyreuther et al.[34] derived a linear relationship between the Mn-valence state ($v_{Mn}$) and the exchange splitting ($\Delta E_{3s}$) of the two Mn-3 s peaks,

$$v_{Mn} = 9.67 - 1.27\Delta E_{3s}/eV. \tag{2}$$

The observed energy splitting ($\Delta E_{3s} = 5.1(5)eV$) gives $v_{Mn} = 3.2(5)$, which is comparable with the expected $v_{Mn} = 3.33$ for the **LP3** compound.

The valence band (**VB**) spectrum of the **LP3** compound was recoded in its paramagnetic state (~ 295 K) and is shown in Fig. 1d. The primary contributions of O-2p, Mn-3d and Pr-4f. orbitals are present in the VB spectrum. The observed valence band offset is ~ 1.1 eV, calculated from the linear regression fit along the leading edge of the valence band spectrum[35,36]. Due to the crystal field effect a distinct feature of the Mn-3d($e_g$) orbital is expected[33,35,37] near ~ 0.5 eV in the ferromagnetic region. However, as this VB spectrum was recorded at room temperature (paramagnetic state) the Mn-3d($e_g$) state is less pronounced. Thus, a low temperature XPS along with theoretical calculations is required to make clear identification of the valence band states in the ferromagnetic region.

**Ion beam analysis (IBA).** The magnetic properties of a manganite oxide are much influenced by small variations of the chemical composition and oxygen stoichiometry. Moreover, impurity phases will have a direct influence on the measured magnetic properties. Non-magnetic impurity phases will have a negative impact on the estimated MCE as it will contribute to the overall mass of the compound. The calculation of the isothermal entropy change and other magnetocaloric properties (described later) are directly related with the magnetic moment per unit mass, it must be confirmed that there are no other phases (even non-magnetic) contributing to the overall mass of the compound. Results from XRPD or XPS measurements are insufficient to extract the chemical composition of a material with good enough accuracy. For example, in La$_{0.7-x}$Pr$_x$Sr$_{0.3}$MnO$_3$ ($x = 0.02$, 0.05, 0.1, 0.2, 0.3), $T_C$ varies from 353.58 K to 295.84 K[38], while the XRPD patterns (except for $x = 0.3$) look very similar and it seems impossible to justify the large change of $T_C$ from the XRPD results. Therefore, IBA has been performed to shed further light on the details of the sample composition.

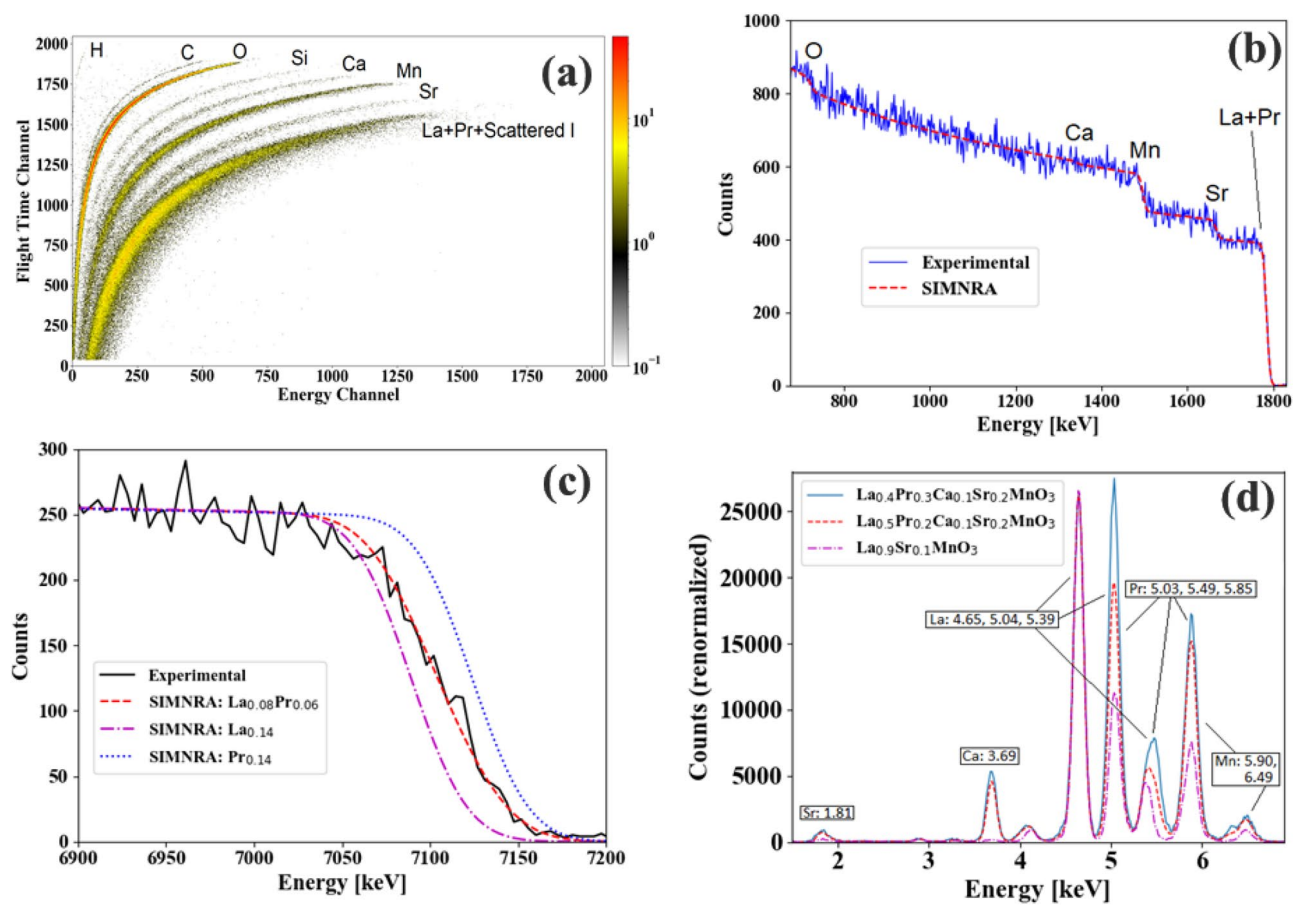

**Figure 2.** (**a**) Raw ToF-ERDA data from sample LP3. (**b**) $^4$He RBS data with overlaid SIMNRA calculated spectrum for sample **LP3**. (**c**) $^{12}$C$^{3+}$ RBS data for the La + Pr signal edge, verifying that the expected ratio reproduces the measured data while calculations using only La or Pr do not. (**d**) PIXE data for sample LP3 and a comparison with spectra presented in Ref.[12]. The spectra have been normalized to the L$_\alpha$-signal from La at 4.65 keV.

Raw ToF-ERDA data is shown in Fig. 2a. In addition to the expected sample constituents; C, Si and H impurities were detected. The signal labelled "Si" may include counts attributable to Al (which might come from the crucible used for heat-treatment), but is processed as Si below. We attribute the observed impurity signals to surface contamination as in IBA no surface cleaning was performed before the measurement. As support of this conclusion, we note that the impurity signals were absent in the XPS results after sputtering the sample surface with Ar-ions. The C, H and Si signals do appear to extend to depths of order 500 nm based on the ToF-ERDA data, but this is primarily an effect of surface roughness, rather than an actual presence of impurities beneath the surface. Furthermore, the detected impurity elements are nonmagnetic and thus their presence (at ≈ 1 atomic %) should not affect any of the magnetic properties.

Depth profiling of the ToF-ERDA data with Potku[39] and integration from depth $1.5 \times 10^{17}$ at/cm$^2$ to $3 \times 10^{18}$ at/cm$^2$ yielded a first estimation of the sample composition while ignoring La and Pr, whose signals are obscured by primary ions scattering into the detector. The $^4$He$^+$ RBS was treated with SIMNRA[40]. Figure 2b shows the experimental data and a calculated spectrum based on a fit of the fractions of Mn, Sr, La and Pr, which improves the quality of the information about the relative amounts of these elements. La and Pr were not separable with $^4$He$^+$ backscattering, so the La/Pr-ratio was fixed to the expected value for the fitting. The full composition of the sample was calculated from the $^4$He-RBS and ToF-ERDA results by taking the ratios of La, Pr and Sr to Mn from RBS and those of H, C, O, Si and Ca to Mn from ToF-ERDA. The resulting atomic concentrations are given in Table 2. The number of counts for which the H, C, O, Si and Ca fractions were calculated from the ToF-ERDA data are approximately 40 (H), 300 (C), 18,000 (O), 400 (Si), 1000 (Ca) and 8000 (Mn). This gives rise to statistical uncertainties, which are treated further below. For H, the main error comes from ion-induced gas release during the measurement as well as an uncertainty in the detection efficiency. A check for ion-induced release of O was performed by dividing the measurement into chronological slices and tracking the time-evolution of the O-signal. No decrease with time was observed, and as such the measurement did not suffer from any noticeable ion-induced release of O. The error here comes rather from the statistical uncertainty in the Mn-signal, which is around 1% and an uncertainty of approximately 5% in the relative detection efficiency. In addition to these error sources, the C, Si and Ca signals include a non-negligible number of background counts. A region of similar size as that giving rise to the Ca signal, but selected between the Ca and Mn regions contains approximately 20–30 counts, i.e. around 5–10% of the C and Si signals and 2–3% of the Ca signal. Combining these error sources yields the error margins in Table 2. For Mn, Sr and La + Pr the relative statistical uncertainties in the number of counts from RBS are 3%, 5% and 1%, respectively. The ratio (La + Pr)/(Sr + Ca) is measured as 2.14 ± 0.21. The error in the fraction of Mn, which is used as a normalization element to combine the RBS and ERDA data, is included here in addition to the explicit errors in the individual elemental concentrations.

| Element | Expected atomic % | Observed atomic % |
|---------|-------------------|-------------------|
| La | 8 | 8.2[a] |
| Pr | 6 | 6.1[a] |
| Ca | 2 | 2.4(2) |
| Sr | 4 | 4.3(2) |
| Mn | 20 | 19.1(6) |
| O | 60 | 57(3) |
| H | 0 | 0.7(4) |
| C | 0 | 1.0(1) |
| Si | 0 | 1.2(1) |

**Table 2.** Atomic percentages of the **LP3** sample as measured by ToF-ERDA (H, C, O, Si, Ca) and RBS (Mn, Sr, La, Pr). [a]The error in the total amount of La + Pr is ± 0.4 at.% (± 3% relative) due to the statistical uncertainty in the number of counts attributed to Mn, while the individual fractions of La and Pr are approximate (see Fig. 2c,d).

In order to verify that the La/Pr ratio is close to the expected value, the RBS data obtained with $^{12}C^{3+}$ was compared to SIMNRA calculations where either only La, only Pr or both La and Pr with the expected ratio were considered for the La + Pr signal. In all three cases, the entire spectrum can be well reproduced, except for the La + Pr signal edge around 7.1 keV, which is reproduced by a La/Pr ratio of 4/3 as shown in Fig. 2c, indicating an actual ratio close to the intended value. Further, a rough estimation of the Pr/La ratio can be made from the PIXE data recorded during the RBS measurement with the $^4He^+$ beam. Figure 2d shows the PIXE spectrum obtained from the $La_{0.4}Pr_{0.3}Ca_{0.1}Sr_{0.2}MnO_3$ sample compared to those displayed in Fig. 4 of Ref. [12]. The combined peak of the $L_\alpha$-signal from Pr and the $L_\beta$-signal from La close to 5 keV increases in intensity relative to the $L_\alpha$-signal from La at 4.65 keV as the Pr concentration is increased. A comparison of the peak areas as outlined in Ref.[12] yields an approximate ratio of Pr to La that is higher for the $La_{0.4}Pr_{0.3}Ca_{0.1}Sr_{0.2}MnO_3$ sample than for the $La_{0.5}Pr_{0.2}Ca_{0.1}Sr_{0.2}MnO_3$ sample by a factor of 1.88, in accordance with the expected value of 1.875.

From the above analysis, it is clear that the **LP3** compound has the desired composition. The effect on magnetic interactions due to the presence of electron–hole pairs is discussed further below. The IBA and XPS results combined confirm the expected electron/hole ratio in the compound and the absence of impurity phase contributions to the overall mass of the sample.

**Magnetic properties.** The temperature dependence of the field cooled magnetization ($M(T)$) using an applied magnetic field of 0.01 T is presented in Fig. 3a. The $M(T)$ curve reveals that the sample exhibits a paramagnetic-ferromagnetic transition at a temperature ($T_C$) close to room temperature. The value of $T_C$ determined from the $\partial M/\partial T$ versus $T$ curve (not shown here) was found to be 289 K, which presents a main factor in room temperature cooling technology[17,41]. A linear temperature dependence of the inverse susceptibility was observed in the paramagnetic region of the **LP3** compound. From the Curie–Weiss fit the extracted Weiss temperature is 291 K, which is close to the value of $T_C$. Moreover, no Griffiths phase like singularity was observed in the **LP3** compound unlike for our previously reported $La_{0.5}Pr_{0.2}Ca_{0.1}Sr_{0.2}MnO_3$ compound[12].

We have studied the magnetic hysteresis by varying the applied magnetic field from − 2 to 2 T at 100 K (well into the ferromagnetic region) and at 285 K (near $T_C$) (Fig. 3c) confirming a negligible hysteresis (5.2 mT and 0.2 mT, respectively) in the studied temperature range. This result represents an advantage as well a reason for the choice of oxide materials as refrigerant materials in MCE based cooling systems compared to giant magnetocaloric materials with a first-order magnetic transition (**FOMT**), which usually show undesirable magnetic and thermal hysteresis making them unsuitable for magnetic refrigeration[7,42,43].

**Magnetocaloric properties:.** We have calculated the magnetic entropy-change from magnetization isotherms using Maxwell's relationship defined as[15],

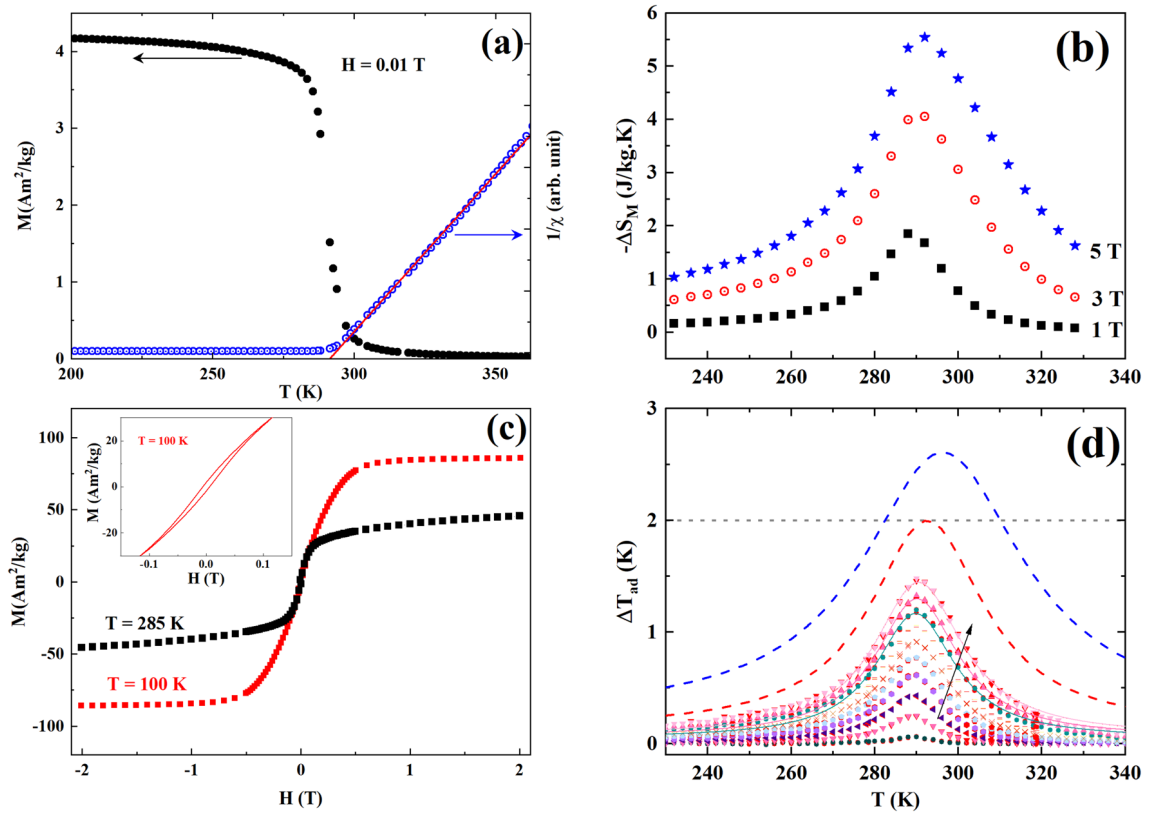

**Figure 3.** (**a**) Magnetization and inverse susceptibility as a function of temperature. (**b**) Isothermal entropy-change as a function of temperature for different $\mu_0 H_f$ values. (**c**) Magnetization versus magnetic field, the inset shows a blow-up of the low field region. (**d**) $\Delta T_{ad}$ versus temperature for different $\mu_0 H_f$ values between 0.1 and 1.9 T in steps of 0.2 T. The dashed upper curves correspond to extrapolated values for $\mu_0 H_f = 3$ K (red) and 5 K (blue). The dotted horizontal line at $\Delta T_{ad} = 2$ K, indicates the minimum requirement for magnetic cooling.

$$\Delta S_M = \mu_0 \int_0^{H_f} \left( \frac{\partial M}{\partial T} \right)_H dH. \tag{3}$$

The magnetic entropy-change as a function of temperature for different $\mu_0 H_f$ values can be seen in Fig. 3b. The value of the isothermal entropy-change increases with increasing magnetic field and achieves its highest value near $T_C$. The maximum $-\Delta S_M$ value was found to be 3.08 J/kg-K under an applied magnetic field of 2 T. This value is large compared to values obtained for some other manganite materials[41,44,45]. It is important to realise the temperature span in which the cooling process will be effective. For this purpose, the RCP is introduced as,

$$RCP = -\Delta S_M^{max} \times \Delta T_{FWHM}, \tag{4}$$

where $-\Delta S_M^{max}$ is the maximum value of isothermal entropy-change and $\Delta T_{FWHM}$ is the full width at half maximum of the isothermal entropy-change curve with respect to temperature. The **LP3** compound shows a high value of RCP, 83.3 J/kg at $\mu_0 H_f = 2$ T.

Figure 3(d) shows the temperature dependence of $\Delta T_{ad}$ measured for different $\mu_0 H_f$ values. The maximum of $\Delta T_{ad}$ is 1.5 K for a magnetic field change of $\mu_0 H_f = 1.9$ T. Also, in the figure we have put the condition $\Delta T_{ad} = 2$ K as a dotted line as Engelbrecht and Bahl[13] have shown that cooling will cease if $\Delta T_{ad}$ drops below 2 K. We have also fitted the $\Delta T_{ad}$ versus temperatures curves using a Lorentzian profile to simulate the response of the material for higher $\mu_0 H_f$ values,

$$\Delta T_{ad} (\mu_0 H_f) = T_0 + \frac{2Aw}{\pi \left( 4(T - T_c)^2 + w^2 \right)}, \tag{5}$$

where $T_0$ is a temperature offset, $A$ is related to the area under the curve and $w$ is the FWHM. Fitting Eq. (5) to the different $\Delta T_{ad} (\mu_0 H_f)$ curves reveal that all parameters ($T_0$, $A$, $w$, $T_C$) are linearly dependent on $\mu_0 H_f$, and for the simulated data at $\mu_0 H_f = 3$ T and $\mu_0 H_f = 5$ T we used parameters from these fits.

Generally, $\Delta T_{ad}$ found in oxide materials is relatively lower compared to that of metallic and intermetallic magnetocaloric materials due to its comparably large specific heat. Near $T_C$, $\Delta T_{ad}$ and $\Delta S_M$ are related by a simple relation[46],

| Compound | $T_C$ (K) | $\mu_0 H_f$ (T) | $-\Delta S_M^{max}$ (J/kg-K) | $\Delta T_{ad}$ (K) | References |
|---|---|---|---|---|---|
| $La_{0.4}Pr_{0.3}Ca_{0.1}Sr_{0.2}MnO_3$ | 289 | 1.9 | 2.98 | 1.5 | This work |
| $La_{0.5}Pr_{0.2}Ca_{0.1}Sr_{0.2}MnO_3$ | 296 | 2 | 1.82 | – | 12 |
| $La_{0.2}Pr_{0.5}Sr_{0.3}MnO_3$ | 299 | 1.8 | 1.95 | 1.2 | 48 |
| $La_{0.4}Pr_{0.3}Sr_{0.3}MnO_3$ | 337 | 1.8 | 1.91 | 1.33 | 48 |
| Gd | 294 | 2 | ~ 5 | ~ 5.8 | 7 |
| $La_{0.6}Ca_{0.4}MnO_3$ | ~ 268 | 0.7 | 1.8 | 0.54 | 14 |
| $La_{0.67}Ca_{0.33}MnO_3$ | 268 | 2.02 | ~ 6.5 | 2.4 | 23 |
| $La_{0.7}Ca_{0.15}Sr_{0.15}MnO_3$ | 338 | 2 | 0.925 | ~ 1.26 | 49 |

**Table 3.** MCE values of magnanite-oxides with $T_C$ near room temperature.

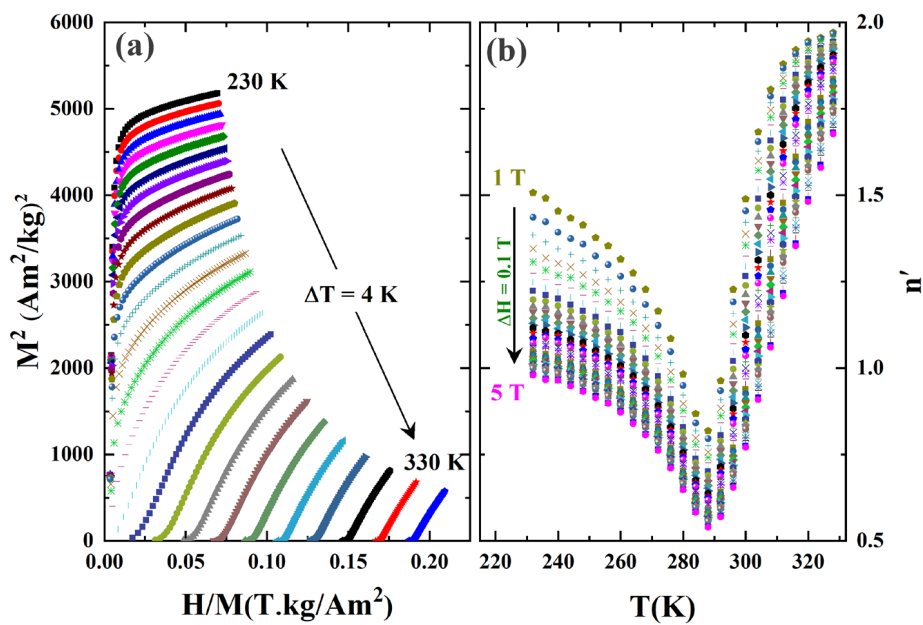

**Figure 4.** (**a**) Arrott plots at different temperatures. (**b**) variation of exponent $n'$ with magnetic field and temperature.

$$\Delta T_{ad} = -\frac{T_C}{C_H(T_C)}\Delta S_M \qquad (6)$$

where $C_H(T_C)$ is the heat capacity at constant magnetic field. Using this relation for the **LP3** compound the estimated heat capacity near $T_C$ is around 574 J/kg-K at a magnetic field of 1.9 T. As a comparison with Gd, the value of $C_H(T_C)$ is ~ 320 J/kg-K at a magnetic field of 1.5 T[47].

Considering the requirement $\Delta T_{ad} > 2$ K, the *RCP* should only be considered between 282 and 310 K, giving $RCP_{\Delta T_{ad} > 2K} = 195.8$ J/kg for $\mu_0 H_f = 5$ T. In Table 3, the values of $-\Delta S_M^{max}$ and $\Delta T_{ad}$ obtained for the **LP3** compound are compared with results obtained for other manganite-oxides having $T_C$ near room temperature.

**Scaling analysis.** The study of critical behaviour provides crucial information of spin interactions in the material. The order of the magnetic transition can be verified using the Banerjee criterion, which is based on Arrott plots ($M^2$ versus $H/M$)[44,45,50]. As shown in Fig. 4a, the positive slopes of the $M^2$ versus $H/M$ curves indicate a second-order phase transition.

However, the Banerjee criterion is based on the mean-field approximation, thus not applicable for all 3D-systems[51]. To avoid confusion relating to the magnetic phase transition, a quantitative criterion of the exponent $n'$ was calculated[51,52], where $n'$ is expressed as,

$$n'(T, \mu_0 H_f) = \frac{dln(|\Delta S_M(T, \mu_0 H_f)|)}{dln(H_f)}. \qquad (7)$$

According to this criterion, for a first-order magnetic transition $n' > 2$ should be observed near $T_C$ while such large values will be absent for a second-order transition. For both types of transitions, the $n'$ value tends to 1 at $T \ll T_C$, and approaches 2 at $T \gg T_C$[51,53]. For small fields the sample will be in a multi-domain magnetic state and Eq. (7)[51] is not applicable, thus applied fields above 1 T have been considered in this calculation. Using the above criterion, from Fig. 4b, a second-order magnetic transition is confirmed for the **LP3** compound.

| Universal class | $\delta$ | $n$ | References |
|---|---|---|---|
| Mean-field (MF) | 3.0 | 0.667 | 57,58 |
| 3D-Heisenberg | 4.8 | 0.627 | 57,58 |
| 3D-Ising | 4.82 | 0.569 | 57,58 |
| Tricritical MF | 5.0 | 0.4 | 57,58 |
| LP3 (low field) | 3.28(3) | 0.750(8) | This work |
| LP3 (high field) | 5.05(3) | 0.598(3) | This work |

**Table 4.** Values of critical exponents.

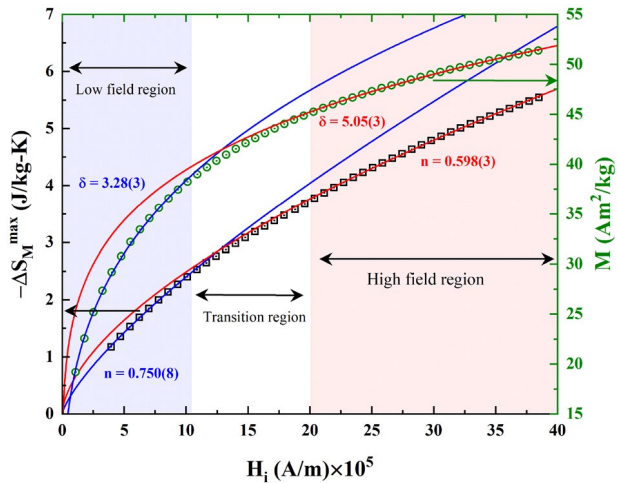

**Figure 5.** Variation of $-\Delta S_M^{max}$ and magnetization with intrinsic field at $T_C$. The red (blue) curves indicate fits to Eqs. (8) and (11) in the high (low) field region.

For a typical second-order phase transition, the critical behaviour of the material is decided by the dimensionality, range of interactions and symmetry of the order parameter[54]. Several universality classes are defined on the basis of the critical exponents $\delta$, $\beta$ and $\gamma$, which are related to the magnetization isotherm at $T_C$, spontaneous magnetization $(M_{sp})$ and initial susceptibility $(\chi)$ by the following relations[54],

$$M(H, T) \propto H^{1/\delta}; \ T = T_C, \tag{8}$$

$$M_{sp}(0, T) \propto (-\varepsilon)^{\beta}; \ T < T_C, \tag{9}$$

$$\frac{1}{\chi}(0, T) \propto \varepsilon^{\gamma}; \ T > T_C, \tag{10}$$

where $\varepsilon = \frac{T - T_C}{T_C}$.

Based on the Arrott-Noakes equation of state, a power law can be derived for the field dependence of $-\Delta S_M^{max}$[54–56] at $T_C$,

$$-\Delta S_M^{max} \propto H_f^{n}, \tag{11}$$

where $n$ depends on the critical exponents as,

$$n = 1 + \frac{\beta - 1}{\beta + \gamma}. \tag{12}$$

In Table 4, the different universality classes are defined on the basis of the critical exponents described above. The magnetization isotherm recorded at $T_C$ and $-\Delta S_M^{max}$ were fitted to the Eqs. (8) and (11), respectively (shown in Fig. 5). From the fitting it is clear that the power law equations are not able to account for the full magnetic field range although the intrinsic field is considered in order to avoid the demagnetization field effect at lower fields[59]. However, in the low field region, the magnetization process proceeds by rapid domain wall movement, which implies that the magnetization does not reflect the intrinsic magnetic properties of the compound. In order to minimize the error in determining the demagnetization factor, D. Kim et al.[59] recommended to use only the data points for which the demagnetization field is less than 50% of the applied field. Following this, for the **LP3** compound applied fields below 0.1 T have been excluded. The fit at lower magnetic fields (intrinsic field range $1.0 \times 10^5$ A/m to $10 \times 10^5$ A/m) resembles the behaviour expected for long-range mean-field like interactions, while at higher magnetic fields (intrinsic field range $20 \times 10^5$ A/m to $39 \times 10^5$ A/m) the obtained values of $n$ and $\delta$ are closer to the values expected for short-range 3D-Ising/Heisenberg and tricritical mean-field interaction models.

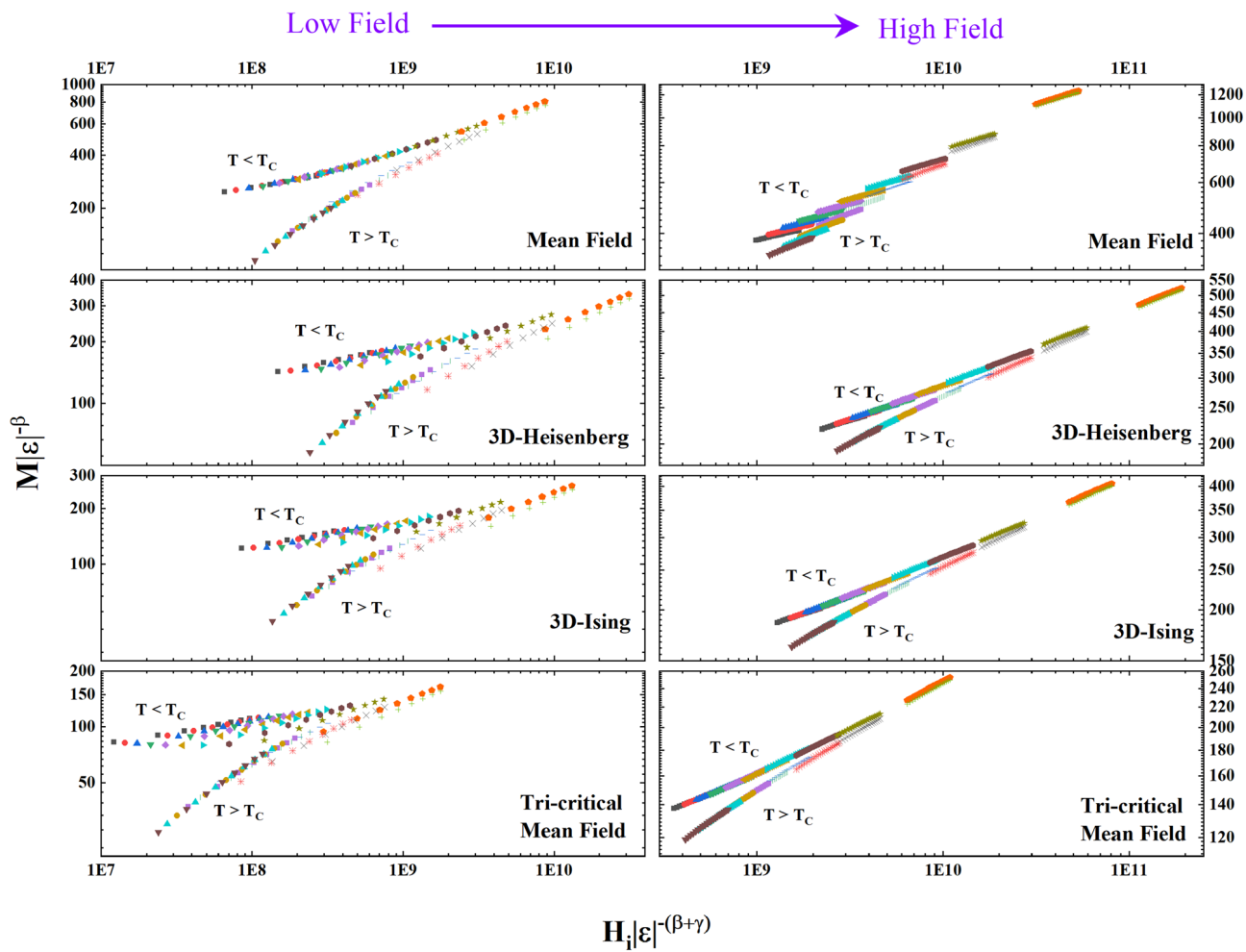

**Figure 6.** $M|\varepsilon|^{-\beta}$ versus $H_i|\varepsilon|^{-(\beta+\gamma)}$. Here $M$ is in Am$^2$/kg and $H_i$ in A/m. A temperature range $T_C \pm 5$ K has been used for the scaling analysis.

For further test of the critical behaviour, the static scaling hypothesis has been considered,

$$M(|\varepsilon|, H)|\varepsilon|^{-\beta} = f_{\pm}\left(H|\varepsilon|^{-(\beta+\gamma)}\right) \tag{13}$$

where the scaling functions $f_+$ and $f_-$ describe positive and negative values of $\varepsilon$, respectively. If suitable values of the exponents $\beta$ and $\gamma$ can be found, then in a $M|\varepsilon|^{-\beta}$ versus $H|\varepsilon|^{-(\beta+\gamma)}$ plot, all data points below (above) $T_C$ will collapse on the same $f_-$ ($f_+$) scaling function[59]. For the **LP3** compound, the scaling hypothesis has been tested using the different universality classes and it was found that different models are needed in the low and high field regions as shown in Fig. 6. In the low field region, it is evident that the mean-field model provides the best fit to the data, while this model fails in the high field region where instead 3D Ising/Heisenberg and tricritical models provide equally good fits.

To correctly identify the magnetic interaction in the high field region, the equation of state described by Arrott and Koakes[50] and reviewed by Kaul[57] has been considered,

$$\left(\frac{H}{M}\right)^{1/\gamma} = a\varepsilon + bM^{1/\beta}, \tag{14}$$

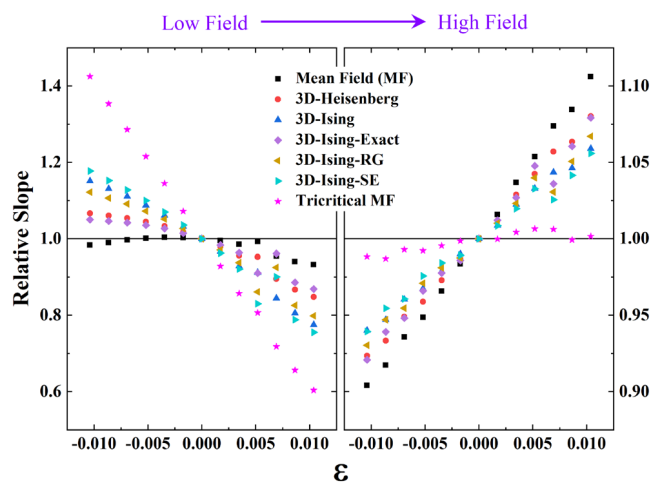

**Figure 7.** Temperature variation of relative slope in modified Arrot plots with respect to the slope of the curve at $T_c$.

where $a$ and $b$ are constants. At $T = T_C$ ($\varepsilon = 0$), a plot of $M^{1/\beta}$ versus $\left(\frac{H}{M}\right)^{1/\gamma}$ should be a straight line passing through the origin if a proper selection of critical exponents has been made. Moreover, the curves corresponding to different temperatures above and below $T_C$ (in the asymptotic critical region[57]) should form parallel lines with respect to the curve at $T_C$. To test this criterion, the relative slopes of the curves normalized with the slope for the $T = T_C$ curve were calculated for the low and high field regions as shown in Fig. 7. Apart from the theoretical models mentioned in Table 4, Z. Zhang et al.[60,61] has studied different 3D-Ising models for a simple orthorhombic system. Among them, the exact solution (3D-Ising-Exact, with $\beta = 0.375$, $\gamma = 1.25$), renormalization group calculation (3D-Ising-RG, with $\beta = 0.340$, $\gamma = 1.244$) and high-temperature series expansion (3D-Ising-SE, with $\beta = 0.312$, $\gamma = 1.25$) are also compared in Fig. 7. From this figure, it is seen that the mean-field model best describes the results in the low field region, while the tricritical mean-field model provides the best fit in the high field region.

It is well known that the $Mn^{3+}$ ion is associated with a large JT distortion owing to the presence of one $e_g$ electron, while for $Mn^{4+}$ there is no JT distortion. As a result, in case of $Mn^{3+}$ lattice sites there are two longer and four relatively shorter Mn–O bonds, in contrast to $Mn^{4+}$ lattice sites that have six nearly equal Mn–O bonds. Thus, in the mixed-valence **LP3** compound the lattice distortions will vary between different Mn-sites. In general, for an isolated $Mn^{3+}$ lattice site there is a large energy barrier for activated hopping of the charge carrier. However, if an electron hops from a distorted $Mn^{3+}$ site to an undistorted $Mn^{4+}$ site, the site that was distorted will become undistorted while the site receiving the electron instead will become distorted. The dynamics of this change occurs rapidly, on the time scale of optical phonons. Since the electron can hop back and forth between the two sites, both sites will effectively have same lattice distortion. The energy barrier for activated hopping of the charge carrier will therefore be much reduced paving the way for ferromagnetic double-exchange interaction. Following Bridges et al.[62], the two Mn sites sharing an electron–hole pair is referred to as a dimeron. The system will thus consist of weakly distorted dimerons and isolated $Mn^{3+}$ lattice site with larger JT distortions. Aggregated dimerons will form clusters where the interactions between Mn moments will be ferromagnetic due to the mobile charge carriers and the double-exchange interaction[24]. At low fields, the magnetization process as well as the critical behaviour will be driven by these clusters and the interaction between clusters will be long-range which explains the mean-field critical behaviour. Following the renormalization group approach of Fisher et al.[63], for a $d$-dimensional system with long-range interactions decaying as $1/r^{d+\sigma}$, the critical exponents will have mean-field values for $\varepsilon = 2\sigma - d < 0$. Thus, for the **LP3** compound at low magnetic field the value of $\sigma$ should be $< 1.5$. Comparably higher fields are required to suppress the large JT distortions associated with $Mn^{3+}$ sites[62] and at the same time to reduce the energy barrier for activated hopping of charge carriers. At high enough field, one can therefore expect charge carriers at such sites to become mobile promoting ferromagnetic double-exchange interaction with neighbouring Mn magnetic moments. This explains the crossover in critical behaviour at higher field and the gradual change from a cluster-like magnetic system to a more homogenous system.

Before motivating the observed tricritical mean-field behaviour in the **LP3** compound, we will consider some perovskite manganite systems which have shown a tricritical mean-field like behaviour. The systems $La_{1-x}Ca_xMnO_3$[64], $La_{0.7}Ca_{0.3-x}Sr_xMnO_3$[38], $La_{0.9}Te_{0.1}MnO_3$[65], $La_{0.5}Ca_{0.5-x}Ag_xMnO_3$[66], $La_{0.5}Ca_{0.5-x}Li_xMnO_3$[67] and $La_{0.7}Ca_{0.3}Mn_{0.91}Ni_{0.09}O_3$[68], have been reported to exhibit a tricritical point neighbouring a first-order phase transition. In particular for the $La_{1-x}Ca_xMnO_3$ system[64], the magnetic transition is first-order for $x = 0.3$, while it is second-order for $x = 0.2$ and 0.4. The critical exponents for the second-ordered compounds resemble those of the tricritical mean-field model at higher fields (at lower fields, the values for $x = 0.2$ is closer to those of the 3D-Ising model). Also, the $x = 0.2$ sample resembles the **LP3** compound with respect to the absence of a Griffiths phase. The parent compound (without Pr-substitution) $La_{0.7}Ca_{0.1}Sr_{0.2}MnO_3$ shows a phase transition governed by 3D-Heisenberg type magnetic interaction[38]. Also, it has been observed that with increasing Pr-substitution the system undergoes a transition from short-range 3D-Heisenberg type interaction to long-range mean-field like interaction in the $La_{0.7-x}Pr_xBa_{0.3}MnO_3$ system[69]. Thus, it could be possible to find a composition $La_{0.7-x}Pr_xCa_{0.1}Sr_{0.2}MnO_3$, which will exhibit a first-order magnetic transition and/or magnetic transition governed by short-range interaction, however this is outside the scope of the present work. To conclude, the **LP3** compound has a second-order magnetic phase transition in the vicinity of a first-order critical point for the $La_{0.7-x}Pr_xCa_{0.1}Sr_{0.2}MnO_3$ system and this results in the tricritical behaviour of the compound at higher magnetic field.

## Conclusions

A detailed analysis of the crystal structure, chemical composition, oxidation state, magnetic and magnetocaloric (direct and indirect) properties of the **LP3** compound has been reported here. An orthorhombic structure was observed from the XRPD patterns, while the chemical stoichiometry and oxidation states were confirmed by the IBA and XPS techniques. A continuous second-order magnetic phase transition was confirmed using both the Banerjee criterion and the quantitative criterion based on the exponent $n'$. Using static scaling analysis and modified Arrott plots, a field induced crossover in critical behaviour was observed. At low magnetic field, the long-range mean-field like interaction dominates due to formation of magnetic clusters consisting of aggregated dimerons, while at higher magnetic field a gradual change from a cluster-like magnetic system to a more homogenous system described by a tricritical behaviour is observed due to the system being close to a first-order critical point.

A transition temperature near room temperature and high values of the isothermal entropy-change and adiabatic temperature change over a wide temperature span (~ 40 K), make **LP3** a potential candidate for magnetic refrigeration near room temperature. For a material to be useful as a magnetocaloric material near room temperature $\Delta T_{ad}$ needs to be larger than 2 K[13], otherwise cooling will cease. In the **LP3** compound, the magnetic field needs to be about 3 T for $\Delta T_{ad} = 2$ K. However, the useful temperature span of the material in a field of 5 T is quite large and estimated to be between 282 and 310 K. Griffith et al.[70] have argued from the temperature averaged entropy-change that materials with large entropy-change over a narrow temperature region are suitable for layered regenerators, while materials with moderate entropy-change over a wide temperature span (as is the case of **LP3**) are suitable for single layer regenerator applications. This makes the **LP3** compound interesting from both fundamental physics and application point of views.

## Acknowledgements
The Swedish Foundation for Strategic Research (SSF, contract EM-16-0039) supporting research on materials for energy applications is gratefully acknowledged. Infrastructural grants by VR-RFI (#2017-00646_9) and SSF (contract RIF14-0053) supporting accelerator operation are gratefully acknowledged. The authors are thankful to Konstantin Skokov for assistance in measuring the adiabatic temperature change, to Vitalii Shtender for recording the XRPD data, and to Sanchari Chakraborti for helping in data plotting.

## Author contributions
The study design was performed by S.G. and R.S. The material was synthesized by R.S. The main part of the manuscript was written by S.G. Direct magnetocaloric measurements and analysis were performed by D.H. I.B.A. measurements and analysis were performed by P.S. Remaining measurements and analysis was performed by S.G. P.S. is responsible for funding and group leader of this work. All authors reviewed and proofread the manuscript.

## Funding
Open Access funding provided by Uppsala University.

## Competing interests
The authors declare no competing interests.

## Additional information
**Correspondence** and requests for materials should be addressed to S.G.

**Reprints and permissions information** is available at www.nature.com/reprints.

**Publisher's note** Springer Nature remains neutral with regard to jurisdictional claims in published maps and institutional affiliations.